# Violation of Fourier's Law and Anomalous Heat Diffusion in Silicon Nanowires

Nuo Yang[1,&], Gang Zhang[2*] and Baowen Li[1,3]

[1]*Department of Physics and Centre for Computational Science and Engineering, National University of Singapore, 117546 Singapore*

[2]*Institute of Microelectronics, 11 Science Park Road, Singapore Science Park II, Singapore 117685, Singapore*

[3]*NUS Graduate School for Integrative Sciences and Engineering, 117456 Singapore*

Email: zhangg@ime.a-star.edu.sg

**Abstract**

We study heat conduction and diffusion in silicon nanowires (SiNWs) systematically by using non-equilibrium molecular dynamics. It is found that the thermal conductivity ($\kappa$) of SiNWs diverges with the length as, $\kappa \propto L^{\beta}$, even when the length is up to 1.1 μm which is much longer than the phonon mean free path. The dependences of $\beta$ on temperature and length are also discussed. Moreover, an anomalous heat diffusion is observed which is believed to be responsible for the length dependent thermal conductivity. Our results provide strong evidence that Fourier's law of heat conduction is not valid in low dimensional nanostructures.

[&]The current address is: Department of Mechanical Engineering, Massachusetts Institute of Technology, 77 Massachusetts Avenue, Cambridge, Massachusetts 02139, USA

Nano materials reveal many excellent mechanical, thermal, and electronic properties. In particular, the thermal conductivity is of great interest because of its strong relevance to the potential application in nanoscale thermoelectric energy generators or on-chip coolers. In macroscopic systems, heat conduction is typically a diffusion process which governed by Fourier's law as:

$$J = -\kappa \nabla T \tag{1}$$

where J is the local heat flux and $\nabla T$ the temperature gradient, $k$ is the thermal conductivity. For bulk material, $k$ is size independent, it depends only on the composite of material. However, whether this is still true or not for nano scale materials is an unsolved question. In fact, by using molecular dynamics (MD) simulations, Zhang and Li demonstrated that the thermal energy transports super diffusively at room temperature in single-walled carbon nanotubes (SWNTs) [1], and the thermal conductivity of SWNT diverges with the length of system as $\kappa \propto L^{\beta}$, where the exponents β depends on the temperature and SWNT diameters, and the value of β is between 0.12 to 0.4 [2] and between 0.11 to 0.32 [3]. Henry and Chen also found an infinite thermal conductivity in 1D polyethylene chains by MD and showed that cross correlations for longitudinal-acoustic phonons are responsible for the divergent thermal conductivity [4, 5]. The breakdown of Fourier's law in nanotube thermal conductivity has also been observed experimentally [6]. These simulations and experimental results have verified the theoretical predictions that Fourier's law is violated for a variety of one-dimensional (1D) lattice models [7-15].

In addition to SWNTs, silicon nanowire (SiNW) is one of the promising nano materials that pushes the miniaturization of microelectronics towards a new level, and has drawn significant attentions because of its reliable structural strength and ideal interface compatibility with the conventional Si-based technology. With the fast and progressive research in experiment and its applications, more and more theoretical efforts have been made to understand the thermal properties of SiNWs. The impacts of diameter, surfaces reconstruction, temperature and isotopic doping have been reported by using MD [16-19], Monte Carlo [20], Boltzmann transport equation [18, 21] and other empirical simulations [22-25]. However, none of the modeling and simulation studies of the thermal property of SiNWs has ever addressed the question whether the Fourier's law of heat conduction is valid in realistic crystalline structures. In this letter, we focus on the length dependence of thermal conductivity of SiNWs by using MD simulation. MD simulations are ideal for thermal conductivity investigation since they can be used to study individual nano materials with realistic crystalline structures. The total thermal conductivity contains contributions from electrons and phonons. In SiNWs, when the carrier concentration is small enough, which is the case for intrinsic and moderately doped silicon, the thermal transport due to electrons is negligible compared to phonon contribution, and thermal conductivity of SiNWs is dominated by phonons. Thus, SiNWs are systems ideally suited for investigations of the validity of Fourier's law in a low dimension. The comprehensive study of thermal conductivities of SiNWs can lead to a deeper understanding of phonon transportation mechanisms in low dimensional systems.

In this letter, we study the thermal conductivity of SiNWs along [100] direction with cross section of 3×3 unit cells (the lattice constant is 0.543 nm), corresponds to cross section area of 2.65 nm$^2$. The length in the longitudinal direction changes from 8 to 2,056 unit cells, which correspond to from 4 nm to 1.1 μm. The system can be readily fabricated by current nanofabrication techniques. The maximum number of atoms in simulation is 147,600.

In our simulations, fixed boundary conditions on the outer surface of NW are adopted, unless otherwise stated, and the non-equilibrium molecular dynamics (NEMD) calculation is performed. To derive the force term, we use Stillinger-Weber (SW) potential for Si [26], which includes both two-body and three-body potential terms. The SW potential has been used widely to study the thermal property of SiNWs and silicon bulk material [18, 27, 28] for its best fit for experimental results on the thermal expansion coefficients.

Generally, in MD simulation, the temperature, $T_{MD}$, is calculated from the kinetic energy of atoms according to the Boltzmann distribution:

$$\langle E \rangle = \sum_1^N \frac{1}{2} m v_i^2 = \frac{3}{2} N k_B T_{MD}, \tag{2}$$

where $\langle E \rangle$ is the mean kinetic energy, $v_i$ the velocity of atom, m the atomic mass, N the number of particles in the system, $k_B$ the Boltzmann constant. However, this equation is valid only at very high temperature ($T>>T_D$, $T_D$ is the Debye temperature). As $T_D = 645$ K for Si, when the system is at room temperature, it is necessary to apply a quantum correction to the MD temperature. Here the average system energy is twice the average kinetic energy based on the equipartition theorem. Then there is an equality of the system energies written in the mechanical and phonon pictures:

$$3Nk_B T_{MD} = \int_0^{\omega_D} D(\omega)\left(\frac{1}{2} + n(\omega,T)\right)\hbar\omega d\omega \tag{3}$$

where $D(\omega)$ is the phonon density of states, $n(\omega,T) = 1/(e^{\hbar\omega/k_B T} - 1)$ is the phonon average occupation number, the ½ term represents the effect of zero point energy and $\omega$ is the phonon frequency, and $\omega_D$ is the Debye frequency. The real temperature $T$ can be deduced from the MD temperature $T_{MD}$ by this relation [27-29]. When $T_{MD}$ is at room temperature, 300K, the real temperature $T$ is 295K, which gives a quite small quantum correction effect on thermal conductivity. In our following study, we focus on the length dependence of thermal conductivity at room temperature and 1000K ( $\gg T_D$ ), respectively.

In order to establish a temperature gradient along the longitudinal direction, the atoms close to the two ends of SiNWs are put into heat baths with temperature **$T_L$** and **$T_R$** for the left and right end, respectively. Nosé-Hoover heat baths [30] are applied. To ensure our results are independent of heat bath, Langevin heat baths [31] are also used. Both types of heat baths give rise to the same results. The expression of the total heat flux along longitude direction is defined as

$$J_l(t) = \sum_i v_{i,l}\varepsilon_i + \frac{1}{2}\sum_{ij}\sum_{i\neq j}r_{ij,l}\left(\vec{F}_{ij}\cdot\vec{v}_i\right) + \sum_{ijk}r_{ij,l}\left(\vec{F}_j(ijk)\cdot\vec{v}_j\right) \qquad (4)$$

where $\varepsilon_i$ is local site energy, $\vec{F}_{ij}$ is two-body force and $\vec{F}_j(ijk)$ is three-body force (details see [19] ).

Simulations are carried out long enough such that the system reaches a stationary state where the local heat flux is a constant along NW and independent of time. After the system reaches a stationary state, a time averaging of the temperature and heat current is performed in 20 ns. In Figure 1 we show the typical time-averaged temperature profile used to compute the thermal conductivity. In the intermediate region, without the thermal boundary resistances effect, the temperature profile is fitted with a linear function and the resulting temperature gradient is used to calculate thermal conductivity as below:

$$\kappa = -\frac{J}{A \times \nabla T}, \tag{5}$$

where $J$ is the heat flux, $A$ is the cross section area, and $\nabla T$ is the temperature gradient.

The dependence of thermal conductivity on the length of SiNW is show in Figure 2. With wire length up to 1.1 μm, the thermal conductivity is less than 40 W/mK, this value is of the same order of magnitude as results obtained from Monte Carlo simulations [20]. The simulation results are higher than the measured thermal conductivity [32] because there is strong phonon scattering effect from rough surface and defects in experiment. It is obvious that with the length increases, the thermal conductivity increases. This demonstrates that the thermal transport in SiNW does not obey Fourier's law, even when the wire length is as long as 1.1 μm. In the description of thermal transport in solids, the phonon mean free path (λ) is an important physical concept. It is a characteristic length scale beyond which phonons scatter and lose their phase coherence. The phonon scattering is highly frequency dependent, which makes the calculation of the mean free path uncertain. For instance, although simple Debye model leads to an estimation of phonon mean free path in silicon at room temperature to be about 41 nm, more detailed dispersion model (a sine function approximation to each acoustic-phonon polarization) show that the mean free path of these phonons that carry heat is about 260 nm [33]. Moreover, using the phonon relaxation time (~10 ps) in SiNWs [18, 34, 35], and the group velocity of phonon as 6400 m/s [33], a value of λ about 60 nm can be achieved. In spite of the large discrepancy in published values for λ (41~260 nm), the maximum SiNW length (1.1 μm) in our calculation is obviously longer than the phonon mean free path of SiNW. In three-dimensional systems, it gives Fourier's law when system scale $L \gg \lambda$. However, our results demonstrate that in SiNWs, Fourier's law is broken even when the length is obviously longer than the mean free path at room temperature. This is consistent with the theoretical prediction and numerical results in low dimensional system [1-5, 7-15] and similar to the experimental results in carbon nanotubes [6].

To understand the underlying mechanism of the divergence of thermal conductivity, we show the phonon density of states spectra (PDOS) in Figure 3. The PDOS spectra are obtained by Fourier transforming velocity autocorrelation function. The PDOS of bulk Si is calculated from a structure of $40\times40\times40$ units$^3$ ($22\times22\times22$ nm$^3$) with periodic boundary condition, which is big enough to obtain the characteristic of bulk Si [34]. The phonon wavelength in a SiNW ranges from the lattice constant to the system size. In the short wavelength side, the shortest phonon wavelength is twice the lattice constant. In the left column of Figure 3, the higher energy part (short wavelength) in PDOS spectra are not sensitive to the wire length. The peak at about 16 THz in PDOS of bulk silicon is appeared even when the NW length is only 4 nm. However, as shown in right column of Figure 3, the lower energy part in PDOS spectra of SiNWs are much different from that of bulk Si. In contrast to the continuous spectra of bulk silicon, there are many discrete peaks in PDOS spectra of NW. For a short SiNW, the low energy phonon density is very small, thus the thermal conductivity is low. With the increase of length, more and more (low energy, long wavelength) phonons are excited, which results in the increase of thermal conductivity. However, due to the size confinement effect, the energy density of lower energy part (acoustic phonon) in PDOS spectra of SiNW are much smaller than that of bulk Si, and leads to the low thermal conductivity of SiNW compared to that of bulk Si.

In Figure 2, we show that the thermal conductivity diverges with the length as $\kappa(L) \propto L^{\beta}$. It should be stressed that the length dependence of thermal conductivity is different in different length regime. At room temperature, when SiNW length is less than about 60 nm, the thermal conductivity increases with the length linearly ($\beta$ =1). For the longer wire (L>60 nm), the diverged exponent $\beta$ reduces to 0.27. The length dependent behavior of harmonic lattice ($\beta$=1) and one dimensional nonlinear lattice model ($\beta$=1/3) [15] are shown in Figure 2 for reference. This critical length, 60 nm is in good agreement with previous predicted value of mean free path, which equals to the phonon relaxation time (~10 ps) [18, 34, 35] times the group velocity of phonon (6400 m/s) [33]. There is weak coupling and interaction among phonons when the length of SiNW is shorter than mean free path. Thus, the phonons transport ballistically, like in harmonic system where no coupling among phonons and thermal conductivity diverged linearly. This result is coincided with the widely accepted result based on kinetic theory [36]. However, when the length of SiNW is longer than mean free path, phonon-phonon scattering dominates the process of phonon transport and the phonon cannot flow ballistically. Our result coincides with previous theoretical results in 1D lattice model in which the thermal conductivity diverged with length with $\beta$ =0.33 when there is strong coupling in the system [13].

In addition, the diverged exponent $\beta$ also depends on temperature. At 1,000K, $\beta$ is only about 0.15 when L>60 nm. The decrease of diverged exponent can be understood as the following. In SiNWs at high temperature, the displacement of atoms will be increased, which will induce more non-linear effects, thus more phonon-phonon interaction, which can decrease the diverged exponent $\beta$ [7].

So far we show the length dependent thermal conductivity of SiNWs with fixed boundary conditions on the outer surfaces. In addition to fixed surface, free boundary condition is a more realistic representation of a wire. However, as the (100) surfaces are not stable [37], we use (110) surfaces to study the thermal conductivity of SiNWs with free boundary condition on transverse direction. The structure is shown in figure 1 in ref. 37. The length dependent thermal conductivity of SiNWs with free boundary conditions is shown in the inset of Fig. 2. It is obvious that thermal conductivity diverges with NW length. This demonstrates the violation of Fourier's law is a general phenomenon in NW, regardless of boundary conditions.

It is worth pointing out that, as is clearly demonstrated by Fig 2. The nanowire with free (transverse) boundary condition has larger value of thermal conductivity than the case with fixed (transverse) boundary condition. The value of divergent exponent, $\beta$, is also larger in the case of free boundary condition. The underlying mechanism is quite clear as the fixed boundary condition causes additional surface scattering which suppresses the thermal conductivity.

Then we study the energy diffusion process in SiNWs. We first thermalize the system to an equilibrium state with temperature $T_0$ (corresponds to energy $E_0$), then atoms in the middle layer (at the center along the longitudinal direction) are given a much higher temperature $T_1$ (higher kinetic energy). The evolution of the energy profile along the chain is then recorded afterwards. Quantitatively, the width of the pulse can be measured by its second moment [1]:

$$\sigma^2(t) = \frac{\sum_i [E_{i,t} - E_0](\vec{r}_{i,t} - \vec{r}_0)^2}{\sum_i [E_{i,t} - E_0]} \quad (6)$$

where $E_{i,t}$ is the energy of atom $i$ at time $t$, $\vec{r}_{i,t}$ is the position of atom $i$ at time $t$, and $\vec{r}_0$ is the position of energy pulse at $t = 0$. To get rid of the influence from reflection, periodic boundary condition along longitudinal direction is used. Then the averaged energy profile spreads as $\langle \sigma^2(t) \rangle \propto t^\alpha$, with $0 \leq \alpha \leq 2$, where <.> means the ensemble average over different realizations. To suppress statistical fluctuations, an ensemble average over 100 realizations is performed. In Figure 4 we show $\langle \sigma^2(t) \rangle$ versus time in double logarithmic scale, so that the slope of the curve gives the value of α. With the NW length of 140 nm, $\alpha = 1.15 \pm 0.01$ and $1.07 \pm 0.01$ at 300K and 1000K, respectively. According to the theory proposed by Li and Wang [12], the exponent of diffusion ($\alpha$) is connected with the exponent of thermal conductivity ($\beta$), $\kappa \propto L^\beta$, by

$$\beta = 2 - 2/\alpha \tag{7}$$

This theory describes the physical connection between energy diffusion and thermal conductivity. For instance, in normal diffusion, $\alpha = 1$, then we have $\beta = 0$, which means that the thermal conductivity is a size-independent constant, this is what we have in bulk material. The phonon transports diffusively, and the Fourier law is valid. However, in another extreme case, such as the ballistic transport, we have $\alpha = 2$, thus $\beta = 1$, namely the thermal conductivity of the system is infinite when the system goes to thermodynamic limit. This is the case for one dimensional harmonic lattice. This is generally believed when the system size is smaller than the mean free path. In another case, if $\alpha < 1$, which we call sub-diffusion case, corresponds to $\beta < 0$, namely, the thermal conductivity of the system goes to zero – the system is an insulator. In the super diffusion regime, namely $\alpha > 1$, we obtain $\beta > 0$, which predicts that the thermal conductivity increases as the length of the system increases. This is what we observed in SiNWs. Using α=1.15 (300K), 1.07 (1,000K) and 1.23 (300K with free transverse boundary), one obtains β=0.26, 0.13 and 0.37 from Eq. (7), which are very close to those values (0.27, 0.15 and 0.4, respectively) calculated directly. Our results demonstrate that the super diffusion is responsible for the length dependent thermal conductivity of SiNWs.

To summarize, we have studied the length dependence of thermal conductivity of SiNWs by using MD simulations. The thermal conductivity increases with NW length up to 1.1 μm. Our results demonstrate that at room temperature, heat conduction in individual SiNW does not obey Fourier's law even though the NW length is much longer than the phonon mean free path. When SiNW length is less than the phonon mean free path, the phonon–phonon interaction can be neglected, then the phonons transport ballistically like in harmonic oscillator, and the thermal conductivity increases with the length linearly. When the length of SiNW is much longer than mean free path, the phonon-phonon scattering plays a key role in the process of phonon transport. However, as already demonstrated in low dimensional lattice model [10-14], phonon-phonon interaction alone is not sufficient to induce a diffusive process, thus the phonons transports super diffusively which results in a diverged thermal conductivity. Our results demonstrate that SiNW is a promising platform to verify phonon transport mechanisms in low dimensional systems.

This work was supported in part by an ARF grant, R-144-000-203-112, from the Ministry of Education of the Republic of Singapore and Grant R-144-000-222-646 from the National University of Singapore.

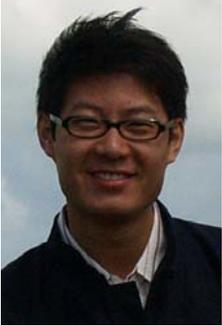

Nuo Yang is a visiting student in the Department of Mechanical Engineering at MIT. He earned his B.S. degree in applied physics from the University of Science & Technology of China (2000), his M.E. in accelerator physics from the Chinese Academy of Science (2003), and a Ph. D candidate in National University of Singapore. His research is focused on thermoelectrics and thermal transportation in low-dimensional structures and thermal interface materials.

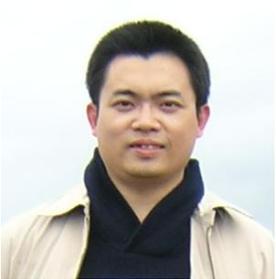

Dr. Gang Zhang is a Senior Research Engineer at the Institute of Microelectronics, Singapore. He graduated in Physics from Tsinghua University, where he also received his Ph.D. in 2002. After his Ph.D., he worked at National University of Singapore (2002-2004). Before joining IME, he was a Post-Doctoral Fellow at Stanford University (2004-2006). His research is focused on the energy transfer and harvesting in nanostructured materials.

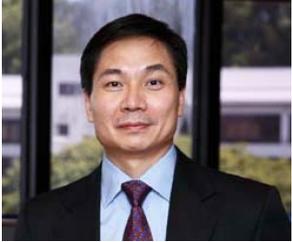

Prof. Baowen Li received his B.Sci (1985) from Nanjing University, M.Sci (1989) from Chinese Academy of Science, Beijing. He obtained Dr.rer.nat degree in 1992 from Universitat of Oldenburg. He joined NUS in 2000 as an assistant professor, and promoted to associate professor in 2003 and full professor in 2007. He is currently Executive Director of NUS Graduate School for Integrative Sciences and Engineering, and Director of Centre for Computational Science and Engineering. His research interests include but not limited to heat conduction in low dimensional systems, complex networks and systems biology, non-equilibrium statistical mechanics, waves propagation and scattering in random/turbulent media etc.

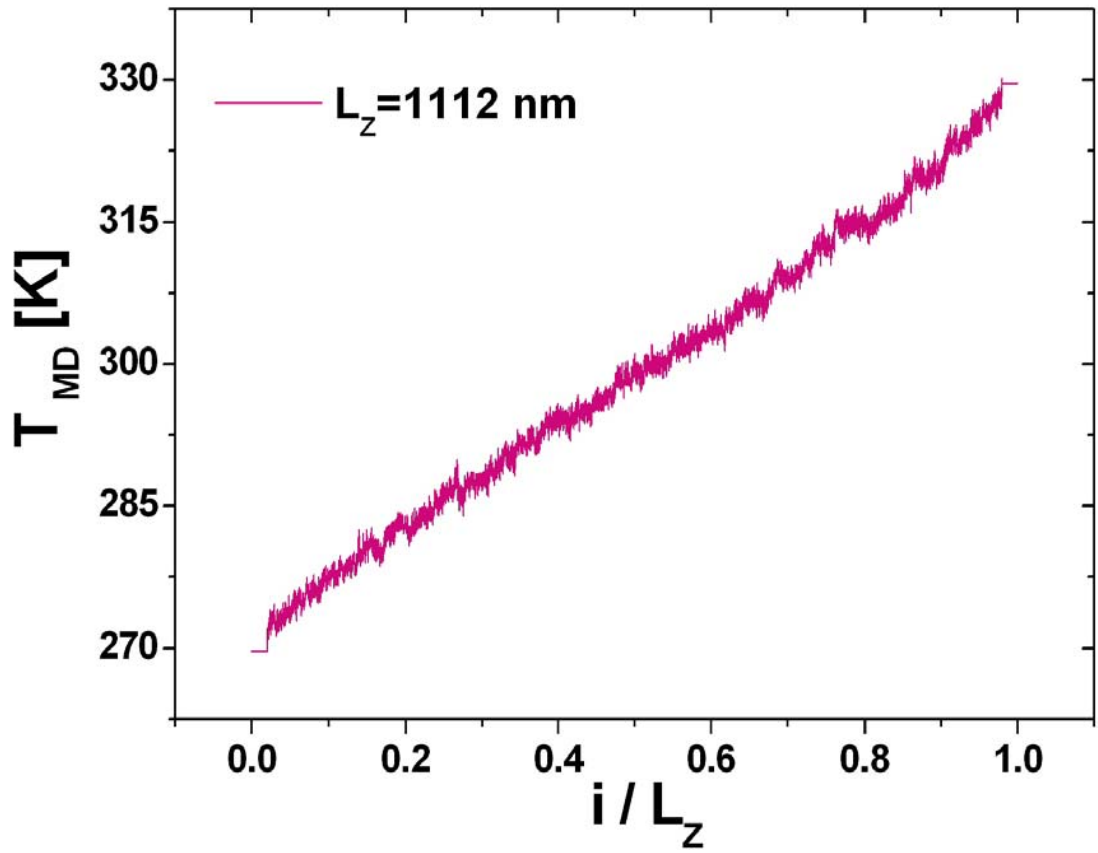

Figure 1. The temperature profile of SiNW with $L = 1.1 \mu m$.

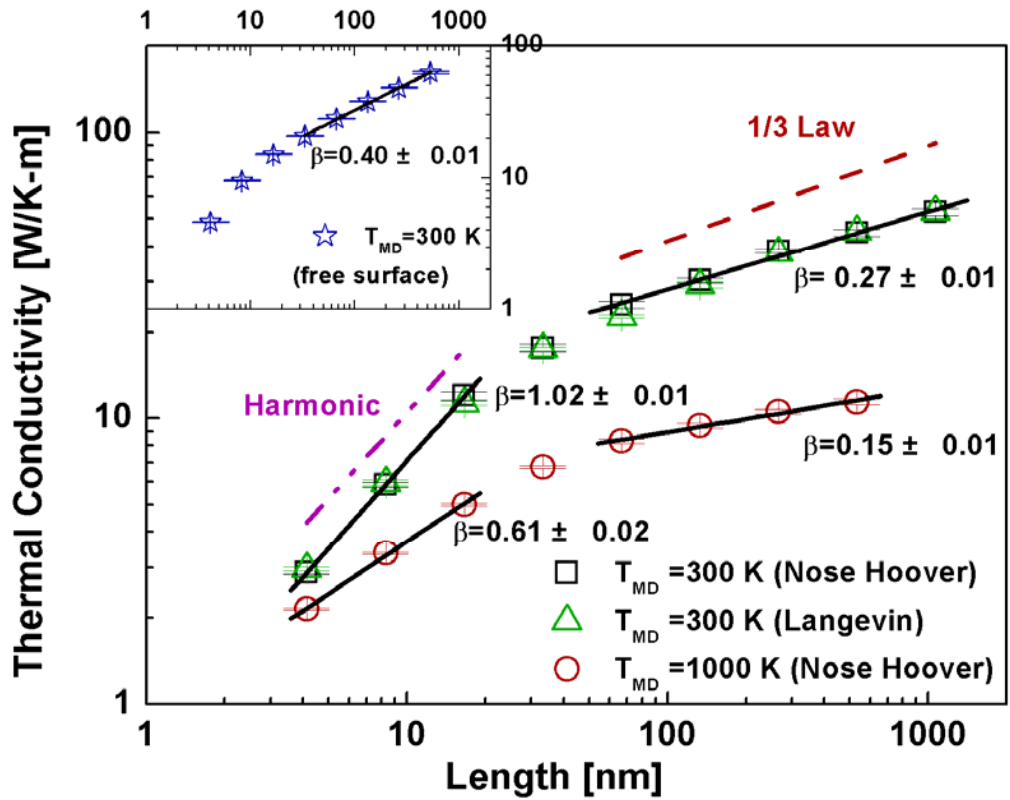

Figure 2. The thermal conductivity of SiNWs (with fixed transverse boundary condition) vs longitude length $L_z$. The results by Nose-Hoover method coincide with those by Langevin methods. The black solid lines are the best fitting ones. The harmonic (dash dotted) and 1/3 (dashed) law are shown for reference. Inset is the length dependent thermal conductivity of SiNWs with free (transverse) boundary condition.

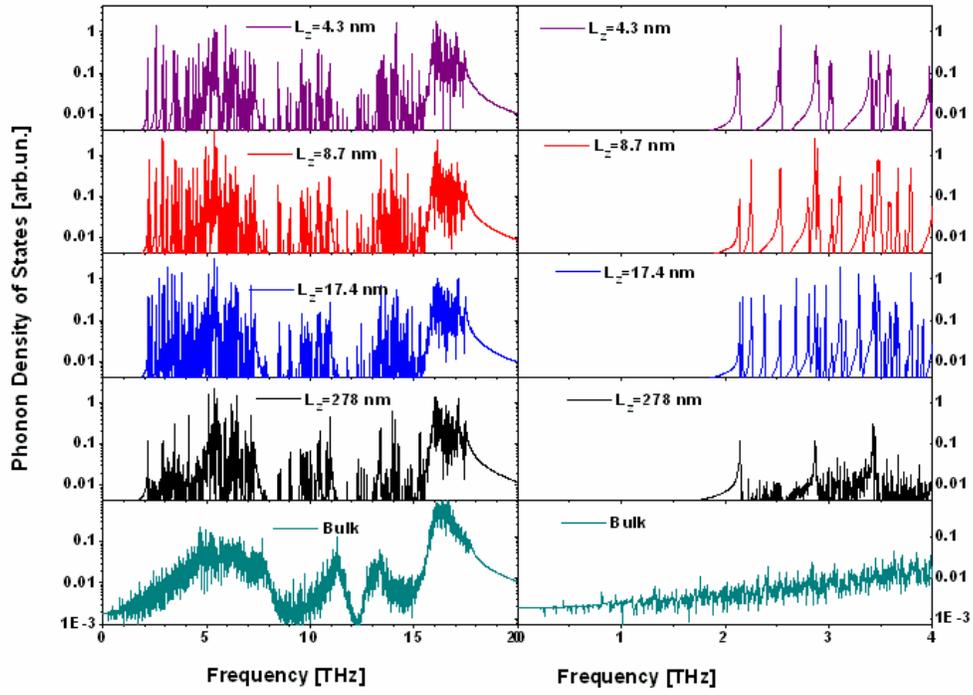

Figure 3. The phonon density of states along the longitude direction of SiNWs with different lengths. The phonon density of states of bulk Si is also shown for reference.

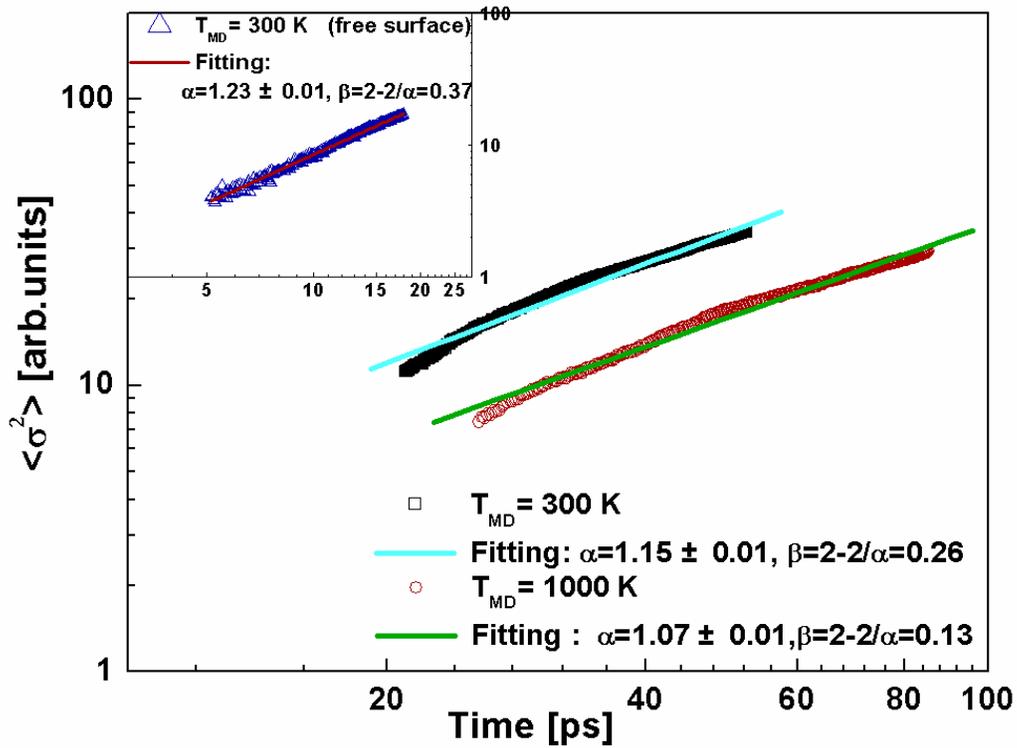

Figure 4. The behavior of energy diffusion in SiNW at room temperature and at 1000K. The length of SiNW is 140 nm. Inset is the behavior of energy diffusion with free (transverse) boundary condition. Please refer to the context for more discussion and comparison of connection between the anomalous diffusion and the anomalous heat conduction.